\documentstyle[12pt,epsf]{article}

\newcommand{\hT}{{\hat T}}
\newcommand{\T}{{\cal T}}
\newcommand{\W}{{\cal W}}
\newcommand{\M}{{\cal M}}
\newcommand{\lan}{{\langle}}
\newcommand{\ran}{{\rangle}}

\newcommand{\vspu}{\vspace*{5mm}}

\newcommand\be{\begin{equation}}
\newcommand\ee{\end{equation}}
\newcommand\bea{\begin{eqnarray}}
\newcommand\eea{\end{eqnarray}}
\newcommand\half{{\textstyle{1\over2}}}

\newcommand{\E}[1]{e^{\textstyle {#1}}}

\makeatletter
\@addtoreset{equation}{section}
\makeatother
\begin{document}
\begin{center}

{\large\bf Green's Functions from Quantum Cluster Algorithms }
\footnote{This work is supported in part by funds provided by the U.S.
Department of Energy (D.O.E.) under cooperative research agreement
DE-FC02-94ER40818.}
\vspu \\
R. Brower$^{\rm a,c}$, S. Chandrasekharan\footnote{ Address after Jan, 1998:
Physics Department, Duke University, Durham NC 27708.}$^{\rm b,c}$ and 
U.-J. Wiese$^{\rm c}$\\
\vspu

$^{\rm a}$ Department of Physics \\
Boston University \\
Boston MA 02215, USA \\
\vspu

$^{\rm a}$ T-8, Theoretical Division \\
Los Alamos National Laboratoy \\
Los Alamos, NM 87545 \\
\vspu

$^{\rm c}$ Center for Theoretical Physics \\
Laboratory for Nuclear Science and Department of Physics \\
Massachusetts Institute of Technology \\
Cambridge MA 02139, USA \\
\vspu
Preprints \\
\ MIT-CTP 2686 \\
\ LA-UR 98-91 \\
\vspu
\end{center}
\vspu
\vspu 

\begin{abstract} \normalsize

We show that cluster algorithms for quantum models have a meaning independent 
of the basis chosen to construct them. Using this idea, we propose a new method
for  measuring with little effort a whole class of Green's functions, once a 
cluster algorithm for the partition function has been constructed. To explain 
the idea, we consider the quantum XY model and compute its two point Green's 
function in various ways, showing that all of them are equivalent. We also 
provide numerical evidence confirming the analytic arguments. Similar 
techniques are applicable to other models. In particular, in the recently 
constructed quantum link models, the new technique allows us to construct 
improved estimators for Wilson loops and may lead to a very precise 
determination of the glueball spectrum.

\end{abstract}
 
 
\newpage

\begin{center}
\noindent{\bf 1. Introduction}
\end{center}

Cluster Algorithms have had a tremendous success in numerically solving some 
interesting field theoretic models \cite{Nie97,Eve97}. Such algorithms have the
potential to drastically reduce the problem of critical slowing down, and 
perhaps even eliminate it completely. The Swendsen-Wang \cite{Swe87} algorithm 
was the first cluster algorithm to be invented.  It has been extended to both 
classical \cite{Uli89} and quantum spin systems \cite{Eve93,Wie94} with 
continuous symmetries. On the other hand, the success of cluster algorithms has
been limited to spin models without frustration. Further, it has not been 
possible to construct an efficient cluster algorithm for Wilson's formulation 
of lattice gauge theory. However, a new formulation of gauge theories, proposed
recently, provides a possible alternative. The corresponding quantum link
models \cite{Cha97} are a quantum version of Wilson's gauge theory, in the same
sense as the quantum Heisenberg model is a quantum version of the 2-d $O(3)$ 
model. A cluster algorithm for the simplest $U(1)$ quantum link model has 
recently been constructed \cite{Bea97}. While in quantum spin models the 
clusters are closed loops, in quantum link models they are world-sheets of 
electric flux strings. It is expected that similar algorithms exist for other 
gauge groups. Thus, it appears that cluster algorithms can also play an 
important role in the numerical solution of gauge theories.

Updating a system with an efficient cluster algorithm accelerates its evolution
through configuration space. However, in quantum models this does not 
automatically imply that one can efficiently measure various Green's functions.
For example, one is interested in creation and annihilation events described by
operators that are not diagonal in the basis in which the path integral is 
constructed.  The measurement of such ``quantum flip'' operators appears to be 
very difficult.  In fact one appears to be restricted to a very limited class 
of Green's functions built from the diagonal operators. For example, let us
consider the quantum XY model expressed in terms of the Pauli matrices
$\sigma^i_x$ at the lattice sites $x$. When we work in the basis where
$\sigma^3_x$ is diagonal, Green's functions for products of $\sigma^3_x$ 
operators can be measured easily.  On the other hand, it appears that Green's 
functions involving $\sigma^1_x$ or $\sigma^2_x$ are difficult --- if not 
impossible --- to measure. A measurement of $\sigma^1_x$ or $\sigma^2_x$ in the
path integral described by eigenstates of $\sigma^3_x$ necessarily ``flips'' 
the state of the corresponding quantum spin. In other words, Green's functions 
of $\sigma^1_x$ or $\sigma^2_x$ receive contributions from superselection 
sectors of the theory, which are not sampled by a conventional Monte Carlo 
algorithm which simply simulates the partition function in a particular basis.

For quantum cluster algorithms, there is a remarkable way out of this impasse. 
By virtue of the auxiliary percolation process used in a quantum cluster Monte 
Carlo algorithm, the requisite information for all quantum Green's functions 
can in fact be extracted.  In this paper, we formulate a method, which we call 
``quantum flip'' measurements, for computing non-trivial Green's functions from
a quantum spin cluster simulation. Our construction of the ``quantum flip'' 
measurement procedure is based on a generalization of the classical 
Fortuin-Kasteleyn \cite{FK} mapping to quantum clusters and on the observation 
that a basis-independent description exists for the resulting quantum random 
cluster model.  For the Swendsen-Wang cluster algorithm, it is well known that 
the Fortuin-Kasteleyn mapping demonstrates the equivalence between the 
partition function of the Ising or Potts model with the partition function of 
an associated random cluster model. This step, which is sometimes ignored in
discussions of cluster algorithms, leads to a purely geometric procedure to 
measure spin correlations. For quantum spin models our generalized 
Fortuin-Kasteleyn map naturally also leads to a systematic and rigorous method 
to measure any operator or Green's function. The quantum Green's functions 
involving operators non-diagonal in a particular cluster update operation are 
related to cuts (or ``flips'') which cleave the original cluster into disjoint 
sub-clusters. Based on this picture, it is possible to apply similar ideas to 
more complicated systems such as gauge theories expressed as quantum link 
models.  In qualitative terms a similar suggestion was made in 
ref.\cite{Pro97}. Our ``quantum flip'' algorithm gives a simple explanation 
of the idea and provides a general approach to measure any Green's function.

The article is organized as follows. In section 2, we reformulate the quantum 
spin model as a model that describes the dynamics of clusters with internal 
quantum degrees of freedom. We use the quantum XY model as an example. The 
reformulation demonstrates the basis-independence of the {\em quantum 
clusters}. In section 3 we show how one can measure any operator using the 
cluster algorithm. In section 4, we illustrate how the observations of sections
2 and 3 are realized in the quantum XY model, by constructing two cluster 
algorithms which are related by a change of basis. We show why the cluster 
properties are independent of the basis, and provide numerical evidence 
confirming the analytic results. In section 5 we suggest how to use similar
ideas in the $U(1)$ quantum link model to obtain improved estimators for Wilson
loops and present our conclusions.

\begin{center}
\noindent{\bf 2. Basis-Independence of Clusters}
\end{center}

In this section, we show that constructing a cluster algorithm for a 
quantum spin model is equivalent to rewriting the path integral in terms of
geometrical objects, which we call {\em quantum clusters}, with internal 
quantum degrees of freedom, in this case quantum spins. We then argue that the 
dynamics within each cluster can be written in a form which is invariant under 
basis transformation of the quantum spin Hilbert space. In the next section, we
discuss how these observations lead to a method for measuring a whole class of 
operators. In order to illustrate the idea we consider the two-dimensional 
quantum XY model. Most of the observations extend trivially to other models 
including quantum link models.

The quantum XY model is described by the Hamilton operator
\begin{equation}
H = \sum_{x,\hat \mu} h_{x,\hat \mu}\;\;;\;\;\;
h_{x,\hat \mu} = \frac{J}{2}\left(
\sigma^1_x\sigma^1_{x+\hat \mu} + \sigma^2_x\sigma^2_{x+\hat \mu} \right).
\end{equation}
Here $\hat \mu = 1,2,...,d$, denotes spatial directions and $x$ denotes the 
position on an $L^d$ lattice. The partition function is 
given by
\begin{equation}
Z = \mbox{Tr} \exp(-\beta H) = \mbox{Tr} \prod_{i=1}^N \exp(-\epsilon H).
\end{equation}
In the last step we have introduced the Suzuki-Trotter discretization such that
$\epsilon N = \beta$. It should be noted that it is not necessary to discretize
time. For quantum systems in a discrete basis the path integral is well-defined
in continuous time, and can be simulated directly in the time-continuum
\cite{Bea96}. In this paper we work at discrete time, but all results have a
direct continuous-time version. In the above form the partition function is not
yet accessible by numerical simulations. In fact, the term $\exp(-\epsilon H)$ 
needs to be expanded further. The basic idea is to write the Hamiltonian as a 
sum of interactions arranged in a checker board pattern such that a minimal 
number of spins interact in a single time step. For example, in $d=2$, one 
decomposes
\begin{equation}
H = H_1 + H_2 + H_3 + H_4.
\end{equation}
This divides the Hamiltonian into four terms in a checker board pattern with
\begin{equation}
H_1 = \sum_{x\in (2m,n)}  h_{x,\hat 1},\;\;\;
H_2 = \sum_{x\in (m,2n)}  h_{x,\hat 2},\;\;\;
H_3 = \sum_{x\in (2m+1,n)}  h_{x,\hat 1},\;\;\;
H_4 = \sum_{x\in (m,2n+1)}  h_{x,\hat 2}.
\end{equation}
Every $H_i$ contains a sum of commuting operators, each of which represents a 
two-spin interaction. Thus $\exp(-\epsilon H_i)$ can be computed easily, and
evolves all the spin states through a single time step. The partition function 
is then defined on a hypercubic lattice with $4N$ time slices. It is equal to 
the trace over the Hilbert space of a product of $2NL^2$ transfer matrices 
connecting pairs of spins. This is because the spins interact in pairs in a 
checker board pattern depending on which of the $\exp(-\epsilon H_i)$ is active
in a particular time step. There are $V/2$ two-spin interactions, where 
$V=4NL^2$ is the volume of the lattice. In the $d=1$ case, one only needs two 
terms in the expansion of the Hamiltonian, i.e., $H = H_1 + H_2$. Then there 
are $2N$ time slices and the partition function is a trace of a product 
constructed from $NL$ transfer matrices, connecting pairs of spins. A typical 
lattice for $d=1$ is shown in figure 1. Generalization to higher dimensions is
straight forward.

The transfer matrix that evolves a pair of spins through a single time step is 
denoted by an operator $T$. Since $T$ represents a two-spin interaction, it is 
equivalent to a $4\times 4$ plaquette matrix. Denoting the basis for the spins 
by $|s\rangle$, the operator
$T$ can be expanded as
\begin{equation}
\label{eq:Trmatop}
T = \sum_{s_1,s_2,s_3,s_4} |s_3 s_4\ran \T(s_1,s_2,s_3,s_4)\lan s_1  s_2 | 
\end{equation}
where $s_1,s_2$ are the spins in one time slice and $s_3,s_4$ are the same 
spins in the next time slice. For example, if we choose the basis to be 
eigenstates of $\sigma^3_x$, the non-zero matrix elements of $T$ are given by
\begin{eqnarray}
\label{eq:XYmatrix}
\T(++++) &= \T(----) &= 1, \\ \nonumber
\; \T(+-+-) &= T(-+-+) &= \cosh(\epsilon J),\; \\ \nonumber
\T(+--+) &= T(-++-) &= \sinh(\epsilon J).
\end{eqnarray}
Here $\pm$ represent the eigenstates of $\sigma^3$ with eigenvalues
$\pm1$.  This leads to the path integral form for the partition function,
\begin{equation}
Z = \mbox{Tr}( \prod_{p=1}^{V/2}T_p) = \sum_{\cal S} \W[s_i] \; ,
\end{equation}
where of course the product over operators $T_p$ for each of the $V/2$
plaquettes must be appropriately ``time'' ordered.  In the last step, we have 
replaced the trace by a sum over spin configurations ${\cal S}$, which is a 
collection of spins $s_i$ located at the space-time points of a 
$(d+1)$-dimensional lattice, weighted by the weight $\W$. The associated 
probability distribution of the spin configuration is given by 
${\cal P}[s_i] = \W[s_i]/Z$. The weight function is a product over the 
four-spin plaquette interactions, which takes the form
\begin{equation}
\W[s_i] =  \T(s_1,s_2,s_3,s_4) \T(s_4,s_5,s_6,s_7) ... 
\T(s_{V-3},s_{V-2},s_{V-1},s_V).
\end{equation}
Due to the {\em trace} structure of the partition function, each spin state 
$s_i$ occurs in two different $\T$ operators. The sum over a 
particular spin $s_i$ makes the partition function independent of the basis 
that is used to describe that spin state. We will see that the 
basis-independence of the partition function extends to the clusters. One can 
rewrite it completely in terms of clusters without referring to the basis of 
the Hilbert space. This observation also leads to similar definitions for the 
expectation values of operators.

We now proceed with the construction of a cluster algorithm and the
Fortuin-Kasteleyn map onto its associated ``random cluster'' model.  The
construction of a cluster algorithm is equivalent to rewriting
$\T(s_1,s_2,s_3,s_4)$ as a sum over products of simpler tensors \cite{Bro89}. 
For example, we can expand
\begin{equation}
\label{eq:cluster}
\T(s_1,s_2,s_3,s_4) = a_0 \M_0(s_1,s_2,s_3,s_4)  +
a_1  \M_1(s_1,s_2,s_3,s_4)  + a_2 \M_2(s_1,s_2,s_3,s_4),
\end{equation}
as a sum of three terms, where we selected a minimal tensor basis,
\begin{equation}
\M_0 = \delta_{s_1,s_3}\delta_{s_2,s_4}, \
\M_1 = \delta_{s_1,s_4}\delta_{s_2,s_3}, \
\M_2 = \sigma^1_{s_1,s_2}\sigma^1_{s_3,s_4},
\end{equation}
with the respective weights uniquely determined to be
\begin{equation}
a_0 = \half (1+\exp(-\epsilon J)), \
a_1 = \half (1-\exp(-\epsilon J)), \
a_2 = \half (\exp(\epsilon J)-1).
\end{equation}
Thus each tensor $\T$ in the partition function is equivalent to a sum of three
tensors, $\M_{n}$, labeled by a plaquette percolation state, $n_p = 0,1,2$.  
This allows us to rewrite the partition function as a sum over $3^{V/2}$ terms 
since there are $V/2$ $\T$s in the partition function.  By choosing one of the 
three terms for each plaquette, one is led to a joint probability distribution,
${\cal P}[s_i;n_p]$, where $n_p$ labels the percolation state of each 
plaquette.  Equivalently, since we have chosen a tensor decomposition in which 
each term is given by a matrix connecting two adjacent spins, each of the
$3^{V/2}$ terms can be 
represented by a graph $G$ with connected components (i.e. clusters) 
${\cal C}_i, i=1,...,N_{G}$. A simple example of a graph in a 1-d model model 
is shown in figure 1.
\begin{figure}[hbt]
\hskip1in
\epsfxsize=100mm
\epsffile{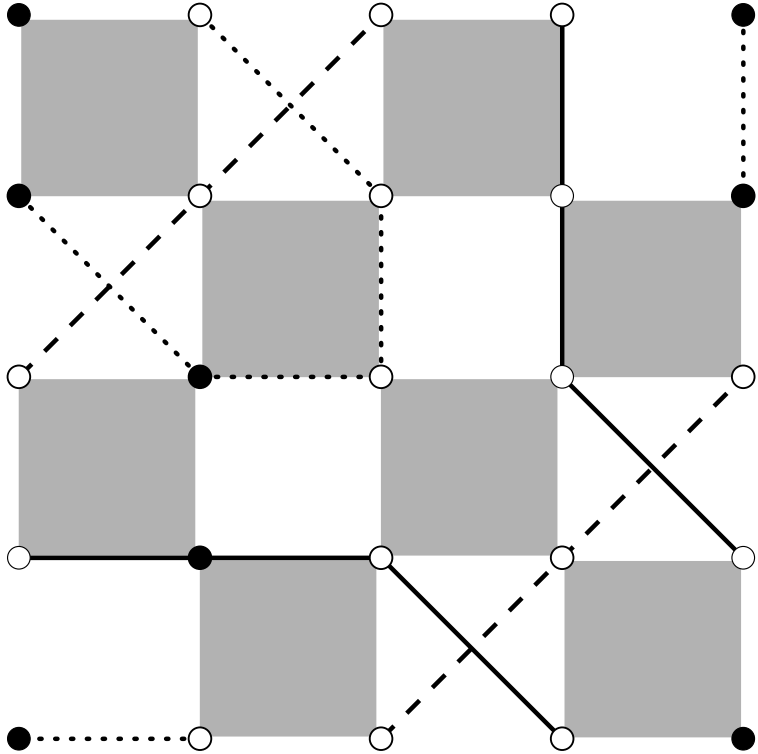}
\caption{\it A typical graph in 1-d with $L=4$, and $N=2$: Since $H=H_1+H_2$, 
one has $2N$ time-slices and $V=2NL$. Every white square represents a two-spin 
interaction. The graph $G$ shown here splits the lattice into three clusters, 
represented by a solid line, a broken line and a dotted line. It is easy to 
read off $w(G)=a^2_0 a^4_1 a^2_2$.}
\end{figure}
 
The Fortuin-Kasteleyn map is based on the simple observation that the joint 
distribution , ${\cal P}[s_i;n_p]$, has two marginal distributions: One,
obtained by summing over the percolation states, is the original spin 
probability distribution,
\be
{\cal P}[s_i] = \sum_{\{n_p\}} {\cal P}[s_i;n_p],
\ee
while the other marginal distribution, obtained by summing over the spin 
variables,
\be
{\cal P}[G] = \sum_{\cal S} {\cal P}[s_i ; \; n_p] \; ,
\ee
is a new ``random cluster model''. The notation on the left-hand side
acknowledges the fact that in our application each distribution of integer 
percolation states $n_p$ uniquely corresponds to a graph $G$.  

It should also be noted that this mapping can be cast in purely operator terms 
for the quantum model. Going back to the operator form of the partition 
function, we have a plaquette operator, $T$, in eq.(\ref{eq:Trmatop}) for each 
four-spin matrix element, $\T(s_1,s_2,s_3,s_4)$. Obviously, the operator $T$ 
can be expressed as a sum of simpler operators, $\sum_n a_n M_n$ with matrix 
elements that reproduce the above expansion eq.(\ref{eq:cluster}) for
$\T(s_1,s_2,s_3,s_4)$.  The contribution to the partition function for a 
particular graph $G$ takes the form of a trace $\mbox{Tr}[M_G]$, where the 
operator $M_G$ is the appropriate time-ordered product of factors $M_{n_p}$ and
the full partition function can be rewritten as
\begin{equation}
Z = \sum_G w(G) \mbox{Tr}[M_G].
\end{equation}
Thus we have defined the probability  distribution,
\be
{\cal P}[G] = \frac{1}{Z} w(G) \mbox{Tr}[M_G],
\ee
for the  associated  quantum random cluster model.
 
The cluster algorithm is a way to realize these dynamics in a Monte Carlo 
method. In a particular basis it reproduces a graph $G$ with the clusters 
${\cal C}_i (i=1,...,N_G$) with the weight $w(G)$ and internal spin 
orientations $\mbox{Tr}[M_G]$. However, since $w(G)$ and $\mbox{Tr}[M_G]$ are 
basis-independent, it is clear that algorithms can be constructed in more than 
one basis. Such algorithms produce clusters with exactly the same geometrical 
properties. We will illustrate this in the case of the quantum XY model in 
section 4. It is possible that the efficiency of algorithms working in 
different bases are different since the cluster growth dynamics would be 
different. Moreover, it is important to realize that in a new basis one may 
choose new cluster decompositions defined by eq.(\ref{eq:cluster}) \cite{Bro89}
which are {\em not} equivalent to a basis change. In this case, the cluster 
properties of two random cluster models are in general inequivalent.

The above picture is a generalization of the well-known Fortuin-Kasteleyn 
mapping \cite{FK} to quantum models. Such a quantum mapping was also considered
in another more technical paper \cite{Kaw95}. Here we present a new way to 
understand the mapping. For example, we have shown that the quantum XY model is
equivalent to a model that involves the dynamics of geometrical objects (in the
present case closed loops), with internal quantum degrees of freedom (in this 
case spins). Summing over the spin degrees of freedom defines a partition 
function that describes the dynamics of the clusters independent of the basis 
used for the spin states. This is at the heart of the Fortuin-Kasteleyn mapping
for the Potts model. As we will show in the next section, the 
basis-independence of the clusters brings into focus some interesting ideas 
on how to measure general Green's functions. In particular, it shows why the 
idea of ref.\cite{Pro97} works and provides a systematic way to measure 
{\em any} operator.

\begin{center}
\noindent{\bf 3. Measurement of Green's Functions}
\end{center}

The basis-independence of the clusters suggests that it must be possible
to measure any operator with a given cluster algorithm. This is indeed 
true. For example, in the XY model it is possible to measure the 
$\sigma^1_x\sigma^1_y$ correlators even when one uses the cluster algorithm in 
the basis where $\sigma^3$ is diagonal. The fact that such measurements are 
possible has recently been suggested in ref.\cite{Pro97}. Based on the above 
ideas, we explain how this is possible. In fact, our approach extends to all 
operators, although the measurement of the above correlator seems to be the 
simplest illustration of our approach.

The expectation value of an operator ${\cal O}_x$, acting at the site $x$,
is defined by
\begin{equation}
\label{eq:Opdef}
\lan {\cal O}_x \ran = \frac{1}{Z} \mbox{Tr} [{\cal O}_x \exp (-\beta H)].
\end{equation}
Rewriting the above expression one obtains
\begin{equation}
\label{eq:Opclust}
\lan {\cal O}_x \ran = \frac{1}{Z}\sum_G w_G \mbox{Tr}[{\cal O}_x M_G].
\end{equation}
Let us consider the steps that lead to the numerator of eq.(\ref{eq:Opclust}). 
The numerator of eq.(\ref{eq:Opdef}) can be written as a product of transfer 
matrices $T$ as described in the previous section. However, now there is an 
additional operator $O_x$ acting on a specific site in a particular time-slice 
$t$. Then there are two transfer matrices adjacent to this site. The matrix 
$T^-$ connects the site with the previous time-slice, and $T^+$ connects it 
with the next time-slice. The operator ${\cal O}_x$ is sandwiched between the 
two. In the next step each $T$ is expanded into three terms and the whole 
expression is again written as a sum of $3^{V/2}$ terms. Each of the $3^{V/2}$ 
terms represents a graph $G$ with clusters ${\cal C}_i$, ($i=1,...,N_G$). The 
weights $w(G)$ are the same as before. However, due to the presence of the 
operator ${\cal O}_x$ the previous expression is modified. It now contains the 
matrix ${\cal O}_x$ appropriately sandwiched between the tensors $\M_0$, 
$\M_1$, or $\M_2$ coming from $T^+$ and $T^-$. In the example of the quantum XY
model, where the clusters are closed loops and $M_G$ is a product of 
$\delta_{s s^\prime}$ or $(\sigma^1)_{s,s^\prime}$, one simply inserts the 
matrix $({\cal O}_x)_{s,s^\prime}$ at the right place. When the cluster turns 
back in time, the appropriate matrix to be inserted is 
$({\cal O}_x^T)_{s,s^\prime}$. Such subtleties must be kept in mind.

It is easy to find the right prescription to measure any operator using the 
cluster algorithm. The cluster algorithm produces the graph $G$ with the 
weight,
\begin{equation}
{\cal P}[G] = \frac{1}{Z} w(G) \mbox{Tr}[M_G].
\end{equation}
Thus we get the formula
\begin{equation}
\label{eq:Opmeas}
\lan {\cal O}_x \ran = 
\lan \frac{\mbox{Tr}[{\cal O}_x M_G]}{\mbox{Tr}[M_G]} \ran.
\end{equation}
To evaluate this expression one must perform all possible cluster flips, which 
can be done very efficiently using an improved estimator. As we will see in the
next section, the above formula can be easily extended to measure any two-point
Green's function in the quantum XY model.

\begin{center}
\noindent{\bf 4. Two-Point Green's functions}
\end{center}

Here we will demonstrate the power of the above ideas by measuring various 
two-point Green's functions in the quantum XY model, using the cluster 
algorithm in the $\sigma^3$ basis. We show how the dynamics of the clusters 
allows us to prove analytically that the $O(2)$ symmetry is maintained. We also
construct a cluster algorithm in a different basis, and show how the same 
features emerge in the new basis.

Let us consider a cluster algorithm obtained by decomposing the transfer matrix
as given in eq.(\ref{eq:cluster}).  The rules for cluster growth are given in 
figure 2. 
\begin{figure}[hbt]
\hskip1in
\epsfxsize=100mm
\epsffile{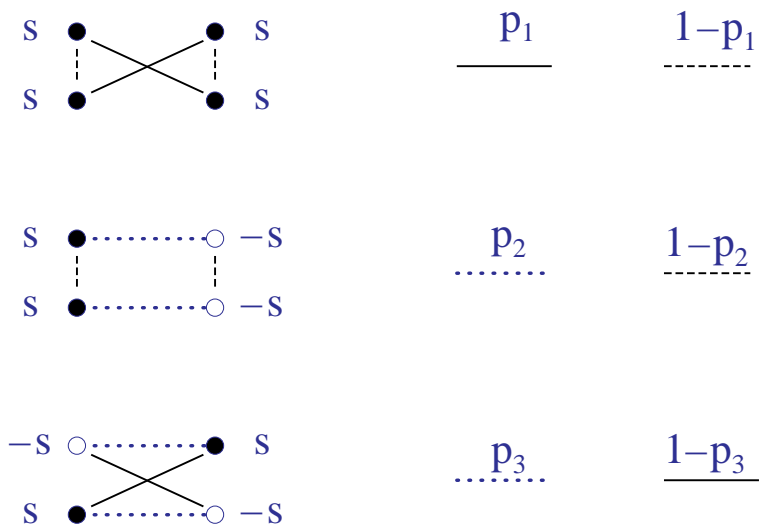}
\caption{\it The rules for creating the bonds that define the graph $G$: The 
sites are described by filled circles and the time axis is vertical. The spin 
states in the $\sigma^3$ basis are denoted by $s$, which can be $\pm$. The 
probabilities for choosing the bonds are $p_1=[1-\exp(-\epsilon J)]/2$, 
$p_2=[\exp(\epsilon J)-1]/[2\cosh(\epsilon J)]$, and 
$p_3=[\exp(\epsilon J)-1]/[2\sinh(\epsilon J)]$.}
\end{figure}
One can easily derive some properties of the clusters.  They are closed loops, 
since each site is connected to two bonds, one for an interaction forward, and
the other for an interaction backward in time. Further, the bonds connect spins
of the same orientation when the cluster grows forward or backward in time.
However, when they connect spins in the same time slice, they connect opposite
spins. Globally, this leads to an interesting property of the clusters. If the 
clusters have positive spins when they grow forward in time, they have negative
spins when they turn back and grow backwards in time. This property of the
clusters is important for proving that the correlators 
$\lan\sigma^1_x\sigma^1_y\ran$ and $\lan\sigma^2_x\sigma^2_y\ran$ are equal.

Let us measure the correlator $\lan\sigma^1_x\sigma^1_y\ran$ with the above 
cluster algorithm. Note that since $\sigma^1_x$ changes the spin state, it is a
priori unclear how to measure $\lan\sigma^1_x\sigma^1_y\ran$ , given just the 
spin configuration. In fact, a spin configuration that contributes to the 
correlator will {\em never} be generated in the cluster algorithm. However, 
eq.(\ref{eq:Opmeas}) tells us to measure 
$\mbox{Tr}[\sigma^1_x\sigma^1_y M_G]/\mbox{Tr}[M_G]$ for each graph generated 
by the cluster algorithm.\footnote{In the interest of simplicity we ignore the 
fact that when the operators $\sigma^1_x$ and $\sigma_y$ are on different 
``time'' slice the product $\sigma^1_x\sigma^1_y M_G$ should, of course, be 
appropriately ``time'' ordered.} Each graph $G$ represents a set of clusters 
${\cal C}_i$ ($i=1,...,N_G$). Since $M_G$ is a product of $\delta_{s s^\prime}$
or $\sigma^1_{s,s^\prime}$, the trace contracts the spins on each cluster 
separately. Thus $\mbox{Tr}[M_G]$ becomes a product of traces for each cluster,
i.e.,
\begin{equation}
\mbox{Tr}[M_G] = \prod_{i=1}^{N_G} \mbox{Tr}[M_{{\cal C}_i}],
\end{equation}
where $N_G$ is the number of clusters in $G$. Further, given one spin in the 
cluster, the orientations of all other spins in the cluster are determined. 
This means that $\mbox{Tr}[M_{{\cal C}_i}] = 2$. In other words,
$\mbox{Tr}[M_G]$ measures the number of configurations obtained from all
possible flips of the clusters. Since each cluster contributes two
configurations, $\mbox{Tr}[M_G] = 2^{N_G}$. On the other hand,
$\mbox{Tr}[\sigma^1_x\sigma^1_y M_G]$ is the same, except that at the sites $x$
and $y$ the spins are now flipped. Clearly, if $x$ and $y$ were in different 
clusters, this flip would be inconsistent going around the closed loop, which 
means that $\mbox{Tr}[\sigma^1_xM_{{\cal C}_i}] = 0$. However, when $x$ and $y$
are both in the same cluster, the contribution is exactly the same as 
$\mbox{Tr}[M_G]$. Thus, each time a cluster is produced that includes both $x$ 
and $y$, we get a contribution 1 to the measurement of
$\lan\sigma^1_x\sigma^1_y\ran$. Due to the translational invariance of the 
Hamiltonian, it is possible to consider any pair of points $x$, $y$ within a 
given cluster and count it as a contribution to the correlator. In this way, it
is possible to obtain the complete two-point correlation function 
$\lan\sigma^1_x\sigma^1_y\ran$.

What about the $\lan \sigma^2_x\sigma^2_y \ran$ correlator? Since the model is 
$O(2)$ invariant, it must be equal to $\lan\sigma^1_x\sigma^1_y\ran$. This is 
indeed the case, although the proof is a bit subtle. While an insertion of 
$\sigma^1$ flips the spin, an insertion of $\sigma^2$ flips the spin along 
with a multiplication by a factor $\pm i$. This means that, when $x$ and $y$ 
are in different clusters, $\mbox{Tr}[\sigma^2_xM_{{\cal C}_i}]$ is zero as 
before. However, if they are in the same cluster, one must figure out what the
factors $\pm i$ contribute. Here we invoke a special property of the clusters
to show that there is always exactly one $i$ and one $-i$ factor coming from 
the two $\sigma^2$ insertions. The point is that if $x$ and $y$ are in the 
branch of the cluster that grows in one particular time direction, the spin 
flips from $+(-)$ to $-(+)$ at the first $\sigma^2$ insertion and from 
$-(+)$ to $+(-)$ at the second insertion. This gives one factor $i$ and one 
$-i$. On the other hand, if $x$ is in a branch growing in a particular time 
direction and $y$ is in a branch growing in the opposite direction, the spin 
flips from $+(-)$ to $-(+)$ at the first and from $+(-)$ to $-(+)$ at the 
second $\sigma^2$ insertion. However, since the cluster is growing backward in
time at one of the insertions, the actual matrix element is that of 
$(\sigma^2)^T$ which again gives one $i$ and one $-i$ factor. Thus, if $x$ and 
$y$ are in the same cluster, the contributions to 
$\lan \sigma^1_x\sigma^1_y \ran$ and $\lan \sigma^2_x\sigma^2_y \ran$ are the
same from each graph generated by the cluster algorithm.

It is straightforward to measure $\lan\sigma^+_x\sigma^-_y\ran$ or
$\lan\sigma^-_x\sigma^+_y\ran$ in the same way, and one gets half of the 
contribution of $\lan \sigma^1_x\sigma^1_y \ran$ for each. Thus, it is possible
to measure any two-point Green's function directly using formula 
eq.(\ref{eq:Opmeas}). In fact, one can easily use it also to measure higher 
$n$-point Green's functions, which, however, get contributions from more than 
one cluster. It is interesting that there is a cluster-notion of ``connected'' 
and ``disconnected'' pieces to the four-point Green's function. The connected 
piece comes entirely from within a cluster, while the disconnected piece comes 
from two separate clusters. 

We now consider a change of basis and show the basis-independence of the
clusters using a concrete example. One can consider a basis change given by
\begin{equation}
| s_x \ran = \sum_{s_x^\prime} 
(\E{-i\sigma^1_x\pi/4})_{s_x s_x^\prime } \; | s_x^\prime \ran.
\end{equation}
This basis change can be applied to each of the spin states. In terms of the
new basis, the transfer matrix given in eq.(\ref{eq:Trmatop}) changes to
\begin{eqnarray}
\label{eq:Trmatop1}
\hT = && \sum_{s_1, s_2, s_3, s_4} \;\;
\sum_{s_1^\prime, s_2^\prime, s_3^\prime, s_4^\prime} \;
(\E{-i\sigma^1\pi/4})_{s_3 s_3^\prime}
	(\E{-i\sigma^1\pi/4})_{s_4 s_4^\prime} 
\nonumber \\
&& |s_3^\prime s_4^\prime\ran \; \T(s_1,s_2,s_3,s_4) \;
\lan s_1^\prime  s_2^\prime | (\E{i\sigma^1\pi/4})_{s_1^\prime s_1} 
	 (\E{i\sigma^1\pi/4})_{s_2^\prime s_2}
\end{eqnarray}

The transformed tensor can be written as
\begin{equation}
\T^\prime(s_1,s_2,s_3,s_4) = a_0 \M^\prime_0(s_1,s_2,s_3,s_4)  +
a_1  \M^\prime_1(s_1,s_2,s_3,s_4)  + a_2 \M^\prime_2(s_1,s_2,s_3,s_4)
\end{equation}
where 
\begin{equation}
\M^\prime_0 = \delta_{s_1,s_3}\delta_{s_2,s_4}, \
\M^\prime_1 = \delta_{s_1,s_4}\delta_{s_2,s_3}, \
\M^\prime_2 = \delta_{s_1,s_2}\delta_{s_3,s_4}.
\end{equation}
Notice that the change of basis has only effected $\M_2$, leaving $\M_0$ and 
$\M_1$ unchanged. It is clear that all the arguments presented in the previous 
section are still applicable here. In fact, one can construct a cluster 
algorithm for the quantum XY model in this basis, which is the same as an 
algorithm for a quantum XZ model in the $\sigma^3$ basis. The cluster rules are
given in figure 3.

\begin{figure}[hbt]
\hskip1in
\epsfxsize=120mm
\epsffile{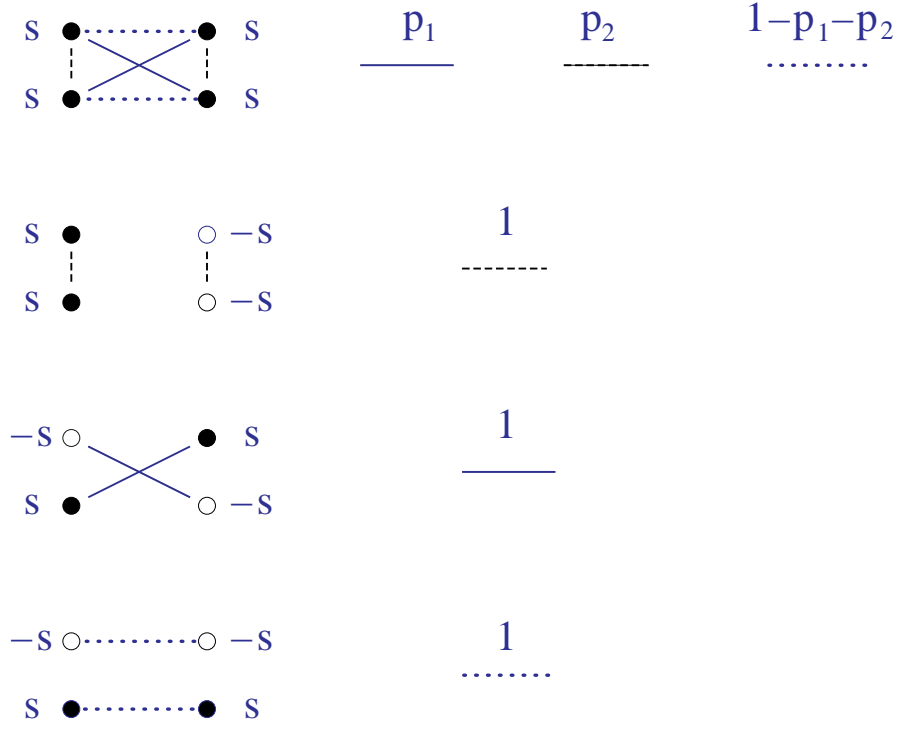}
\caption{\it The rules for creating the bonds that define a graph $G$ in the XY
model in the basis in which $\sigma^2$ is diagonal: The sites are described by 
filled circles and the time axis is vertical. The spin states in the $\sigma^2$
basis are denoted by $s$, which takes values $\pm$. The probabilities for
choosing the bonds are given by
$p_1=[1-\exp(-\epsilon J)]/[\exp(\epsilon J)+1]$ and 
$p_2=[1+\exp(-\epsilon J)]/[\exp(\epsilon J)+1]$.}
\end{figure}

Also in the new basis the clusters are closed loops. However, now all the spins
in a given cluster have the same orientation. Since now we are dealing with the
XZ model, the $\lan\sigma^3_x\sigma^3_y\ran$ correlator must be the same as the
$\lan\sigma^2_x\sigma^2_y\ran$ correlator in the XY model. It is easy to see 
that, if $x$ and $y$ belong to different clusters, 
$\mbox{Tr}[\sigma^3_xM_{{\cal C}_i}]$ is zero. However, when they belong to the
same cluster, since all spins within a cluster have the same orientation,
$\mbox{Tr}[\sigma^3_x\sigma^3_yM_{{\cal C}_i}] = \mbox{Tr}[M_{{\cal C}_i}]$. 
Thus, one gets a contribution 1 to $\lan\sigma^3_x\sigma^3_y\ran$. This is 
exactly what we had in the previous basis for $\lan\sigma^2_x\sigma^2_y\ran$. 
Note that in the previous basis one cannot show that 
$\lan\sigma^3_x\sigma^3_y\ran$ gets a contribution 1 whenever $x$ and $y$ 
belong to the same cluster, since opposite spins can be present in the same 
cluster.

We have obtained numerical results that confirm the above conclusions. We have
updated a $10 \times 10$ lattice with $N=100$ at $\beta J= 1.0$ using both 
the XY and the XZ algorithm. The above arguments imply that both algorithms 
must yield identical cluster properties. The cluster size distribution has 
been measured with a single cluster algorithm. Figure 4 shows the histograms 
for the cluster size distributions obtained with the two algorithms. We see 
that they are identical within statistical errors. In table 1 we give the 
numerical values for the first few moments of the cluster size distributions 
of figure 4. 

\begin{figure}[hbt]
\hskip1in
\epsfxsize=120mm
\epsffile{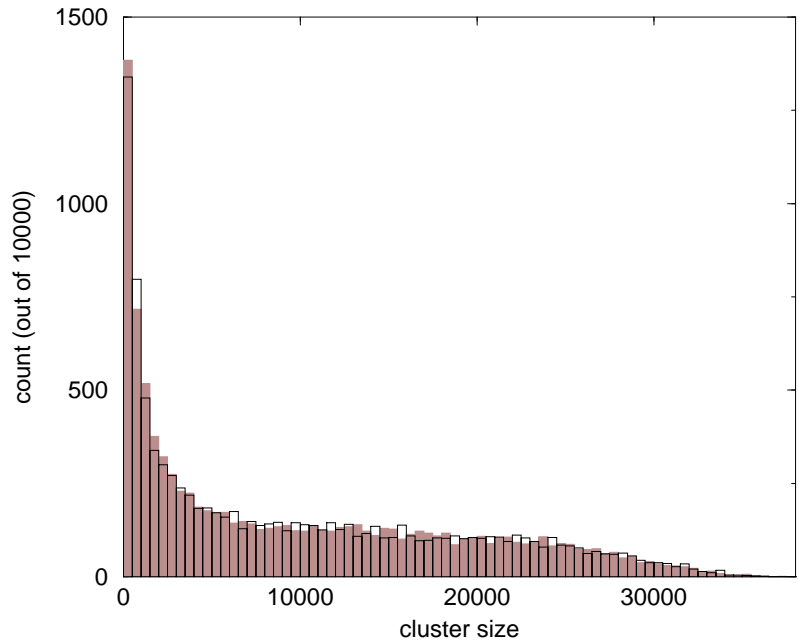}
\caption{\it The histograms of cluster size distribution for the two algorithms
discussed in the text: Both algorithms were run with single cluster updates. 
The data sets contain $10000$ clusters each. The results of the XY algorithm 
are given by the shaded and that of the XZ algorithm are given by the unshaded 
histogram.}
\end{figure}
\begin{table}[hbt]
\begin{center}
\begin{tabular}{|c|c|c|c|}
\hline
Algorithm & $\lan {\cal C} \ran$ & $\lan {\cal C}^2 \ran$ & $\lan {\cal C}^3 
\ran$ \\
\hline
XY & $9.750(120)\times 10^3$ & $1.790(33)\times 10^8$ & 
$3.941(96)\times 10^{12}$ \\
\hline
XZ & $9.818(119)\times 10^3$ & $1.807(34)\times 10^8$ & 
$3.990(95)\times 10^{12}$  \\
\hline
\end{tabular}
\caption{\it The first few moments of the cluster size for the two algorithms: 
The simulations were performed on a $10\times 10$ lattice at $\beta J=1.0$ and 
$N=100$. This clearly demonstrates the basis-independence of the clusters.}
\end{center}
\end{table}

\begin{center}
{\bf 5. Conclusions}
\end{center}

In this article we have developed a new method to evaluate any Green's function
in quantum spin models using cluster algorithms. The method is based on the 
fact that clusters have a meaning independent of the basis. In other words, 
the construction of a cluster algorithm implies that one can rewrite the 
quantum spin model as a random cluster model in a basis-independent way. This 
is well-known in the case of the Potts model using the Fortuin-Kasteleyn 
mapping \cite{FK}. Here we have presented a simple argument showing that such a
mapping also exists for quantum spin models. This mapping provides a new method
to measure any operator once a cluster algorithm has been constructed. In this 
method a measurement of a certain operators reduces to the computation of 
traces in the Hilbert space for each individual cluster. In the quantum XY 
model, this turns out to be quite simple. The method applies to other models as
well, and is useful for measuring quantities which could not be measured using 
standard methods.

  The insight gained here extends easily to quantum link models, which were 
referred to in the introduction. Since these models are gauge invariant, the 
operators of interest are closed Wilson loops. It seems difficult to construct
cluster algorithms in the basis where the Wilson loops can be measured easily. 
For example, in the $U(1)$ quantum link model, a cluster algorithm has been 
constructed in an electric flux basis. The clusters resemble world-sheets of 
electric flux strings. When Gauss' law is imposed, the flux strings form
closed loops and hence sweep out closed world-sheets. Although the Wilson loop 
operators are not diagonal in the electric flux basis, they can be computed 
using the method proposed in this article. The computation requires 
understanding the topology of the closed world-sheets generated by the Monte 
Carlo algorithm. For example, only if the cluster surface separates into two 
pieces when it is cut along the links belonging to the Wilson loop, there is a 
non-zero contribution to the operator. It is easy to show that Wilson loops get
contributions from only one cluster at a time. At present, this is used to
investigate the physics of the $U(1)$ quantum link model. The results of this 
study will be published elsewhere \cite{Bea97}.

The connection with the random cluster model should be understood in more 
detail in the context of quantum cluster algorithms. One intriguing possibility
is a solution of the negative sign problem, at least in a class of models. 
Perhaps the random cluster model suggests a natural basis to avoid negative 
signs. The new insight may also serve as a guide for constructing cluster 
algorithms for quantum link models with non-Abelian gauge groups.

\begin{center}
{\bf Acknowledgement}
\end{center}

One of the authors (S.C.) would like to thank the high energy theory
group at the Fermi National Accelarator Laboratory, where part of this 
research was done, for its hospitality.

\end{document}